\documentclass[AMA,STIX1COL,doublespace]{WileyNJD-v2}

\usepackage{csquotes}
\usepackage{cleveref}
\usepackage{tabularx}
\usepackage{booktabs}
\usepackage{tikz}

\usepackage{enumitem}
\setlist[itemize,1]{itemindent=0.55cm}

\usepackage{xspace}
\newcommand{\ie}{i.\,e.\@\xspace}
\newcommand{\eg}{e.\,g.\@\xspace}

\newcommand*{\ci}[2]{%
\begin{tikzpicture}[scale=0.30]%
    \def\radius{0.5}
    \foreach \x in {1, ..., #1}{
        \fill[fill opacity=0.5,fill=blue] (\x, 0) -- ++(90:\radius) arc (90:-270:\radius) -- cycle;%
    }
    \foreach \x in {1, ..., #2}{
        \draw[thick, black] (\x, 0) circle (\radius);
    }
\end{tikzpicture}%
}%

\articletype{Article Type}%

\received{<day> <Month>, <year>}
\revised{<day> <Month>, <year>}
\accepted{<day> <Month>, <year>}

\begin{document}

\title{Evaluation of tools for describing, reproducing and reusing scientific workflows}

\author[bam]{Philipp Diercks*}
\author[iws]{Dennis Gläser}
\author[dlr]{Ontje Lünsdorf}
\author[kit]{Michael Selzer}
\author[iws]{Bernd Flemisch}
\author[bam]{Jörg F. Unger}

\authormark{Diercks \textsc{et al}}

\address[bam]{\orgdiv{Department 7.7 Modeling and Simulation}, \orgname{Bundesanstalt für Materialforschung und -prüfung (BAM)}, \orgaddress{Unter den Eichen 87, 12205 \state{Berlin}, \country{Germany}}}
\address[iws]{\orgdiv{Lehrstuhl für Wasser- und Umweltsystemmodellierung}, \orgname{Institut für Hydromechanik und Hydrosystemmodellierung, Universität Stuttgart}, \orgaddress{Pfaffenwaldring 61, 70569 \state{Stuttgart}, \country{Germany}}}
\address[kit]{\orgdiv{Institut für Angewandte Materialien-MMS}, \orgname{Karlsruher Institut für Technologie}, \orgaddress{Strasse am Forum 7, 76131 \state{Karlsruhe}, \country{Germany}}}
\address[dlr]{\orgdiv{Institut für Vernetzte Energiesysteme}, \orgname{Deutsches Zentrum für Luft- und Raumfahrt}, \orgaddress{Carl-von-Ossietzky-Straße 15, 26129 \state{Oldenburg}, \country{Germany}}}

\corres{*Philipp Diercks, \orgaddress{Unter den Eichen 87, 12205 \state{Berlin}, \country{Germany}}, \email{philipp.diercks@bam.de}}

\presentaddress{\orgaddress{Unter den Eichen 87, 12205 \state{Berlin}, \country{Germany}}}

\abstract[Abstract]{%
In the field of computational science and engineering, workflows often entail the
application of various software, for instance, for simulation or pre- and postprocessing.
Typically, these components have to be combined in arbitrarily complex workflows
to address a specific research question.
In order for peer researchers to understand, reproduce and (re)use the findings
of a scientific publication, several challenges have to be addressed.
For instance, the employed workflow has to be automated and information on all used 
software must be available for a reproduction of the results.
Moreover, the results must be traceable and the workflow documented 
and readable to allow for external verification and greater trust.

In this paper, existing workflow management systems~(WfMSs) are discussed regarding
their suitability for describing, reproducing and reusing scientific workflows.
To this end, a set of general requirements for WfMSs were deduced from user stories
that we deem relevant in the domain of computational science and engineering.
On the basis of an exemplary workflow implementation, 
publicly hosted at GitHub~(\url{https://github.com/BAMresearch/NFDI4IngScientificWorkflowRequirements}),
a selection of different WfMSs is compared with respect to these
requirements, to support fellow scientists in identifying the WfMSs that
best suit their requirements.
}

\keywords{FAIR, reproducibility, scientific workflows, tool comparison, workflow management}

\jnlcitation{\cname{%
\author{P. Diercks}, 
\author{D. Gl\"aser}, 
\author{O. L\"unsdorf},
\author{M. Selzer},
\author{B. Flemisch}, and 
\author{J. F. Unger}} (\cyear{2022}), 
\ctitle{Evaluation of tools for describing, reproducing and reusing scientific workflows}, \cjournal{International Journal for Numerical Methods in Engineering}, \cvol{2022;XX:X-X}.}

\maketitle

\footnotetext{\textbf{Abbreviations:} API, application programming interface; CLI, command line interface; CPU, central processing unit; DAG, directed acyclic graph; DSL, domain specific language; FAIR, findable accessible interoperable reusable; GUI, graphical user interface; GWL, Guix Workflow language; HPC, high-performance computing; IFDS, internet of FAIR data and services; IRI, internationalized resource identifiers; MPI, message passing interface; SSH, secure shell protocol; WfMS, workflow management system; YAML, YAML ain't markup language}

\section{Introduction}%
\label{sec:introduction}
With increasing volume, complexity and creation speed of scholarly data, humans rely more and more on computational support in processing this data.
The \enquote{FAIR guiding principles for scientific data management and stewardship}~\cite{FAIR2016} were introduced in order to improve the
ability of machines to automatically find and use that data.
FAIR comprises the four foundational principles \enquote{that all research objects should be 
\textit{F}indable, \textit{A}ccessible, \textit{I}nteroperable and \textit{R}eusable (FAIR) both for machines and for people}.
In giving abstract, high-level and domain-independent guidelines, the authors answer the question of what constitutes good data management.
However, the implementation of these guidelines is still in its infancy, with many challenges not yet identified and some of which may not have readily available solutions~\cite{MonsEtAl2020}.
Furthermore, efforts are made towards an Internet of FAIR Data and Services (IFDS)~\cite{EOSC}, which requires not only the data, but also the tools and (compute) services to be FAIR\@.

Data processing is usually not a single task, but in general (and in particular for computational simulations) relies on a 
chain of tools.
Thus, to achieve transparency, adaptability and reproducibility of (computational) research, the FAIR principles must be applied to all components of the research process.
This includes the tools (\ie \textit{any} research software) used to analyze the data, but also the scientific workflow itself which describes how the various processes depend on each other.
In a community-driven effort, the FAIR principles are applied to research software and are extended to its specific characteristics by the FAIR for Research Software Working Group~\cite{fair4rs}.
For a discussion of how the FAIR principles should apply to workflows and workflow management systems~(WfMSs) we refer to the work by Goble~et.~al.~\cite{GobleEtAl2020}.

In addition, in recent years there has been a tremendous development of different tools
(see \eg~\url{https://github.com/pditommaso/awesome-pipeline}) that aid the definition and automation of computational workflows.
These WfMSs have great potential in supporting the goal above which is further discussed in~\cref{sec:introduction_to_workflow_management_systems}.

In this paper, we would like to discuss how WfMSs can contribute to the transparency, adaptability and reproducibility of computational research,
which are aspects that ultimately increase the credibility of research results.
Based on the authors' experience, user stories that are relevant in the domain of computational science and engineering are defined. These
user stories are then used to extract a set of general requirements for WfMSs. Several different tools are compared with respect to these requirements
to support fellow scientists in identifying the tools that best suit their requirements. The list of tools selected for comparison is subjective and certainly not 
complete. However, a GitHub repository~\cite{NFDI4Ing_Scientific_Workflow_2022} providing an implementation of an exemplary workflow for all tools and a short documentation with a link to further
information was created, with the aim to continuously add more tools in the future.
Furthermore, by demonstrating how the different tools could be used, we would like to encourage people to use WfMSs in their daily work.

\subsection{Introduction to workflow management systems}%
\label{sec:introduction_to_workflow_management_systems}

In this paper, we use the term \textit{process} to describe a computation, that is, the execution of a program to produce output data from input data. A process can be arbitrarily complex, but from the point of view of the workflow,
it is a single, indivisible step.
A \textit{workflow} describes how individual processes relate to each other.
Software-driven scientific workflows are often characterized by a complex interplay of various pieces of software executed in a particular order.
The output of one process may serve as input to a subsequent process, which requires them to be executed sequentially with a proper mapping of outputs to inputs.
Other computations are independent of each other and can be executed in parallel.
Thus, one of the main tasks of WfMSs is the proper and efficient scheduling of the individual processes.

\begin{figure}[htb]
    \centering
    \includegraphics[width=.5\textwidth]{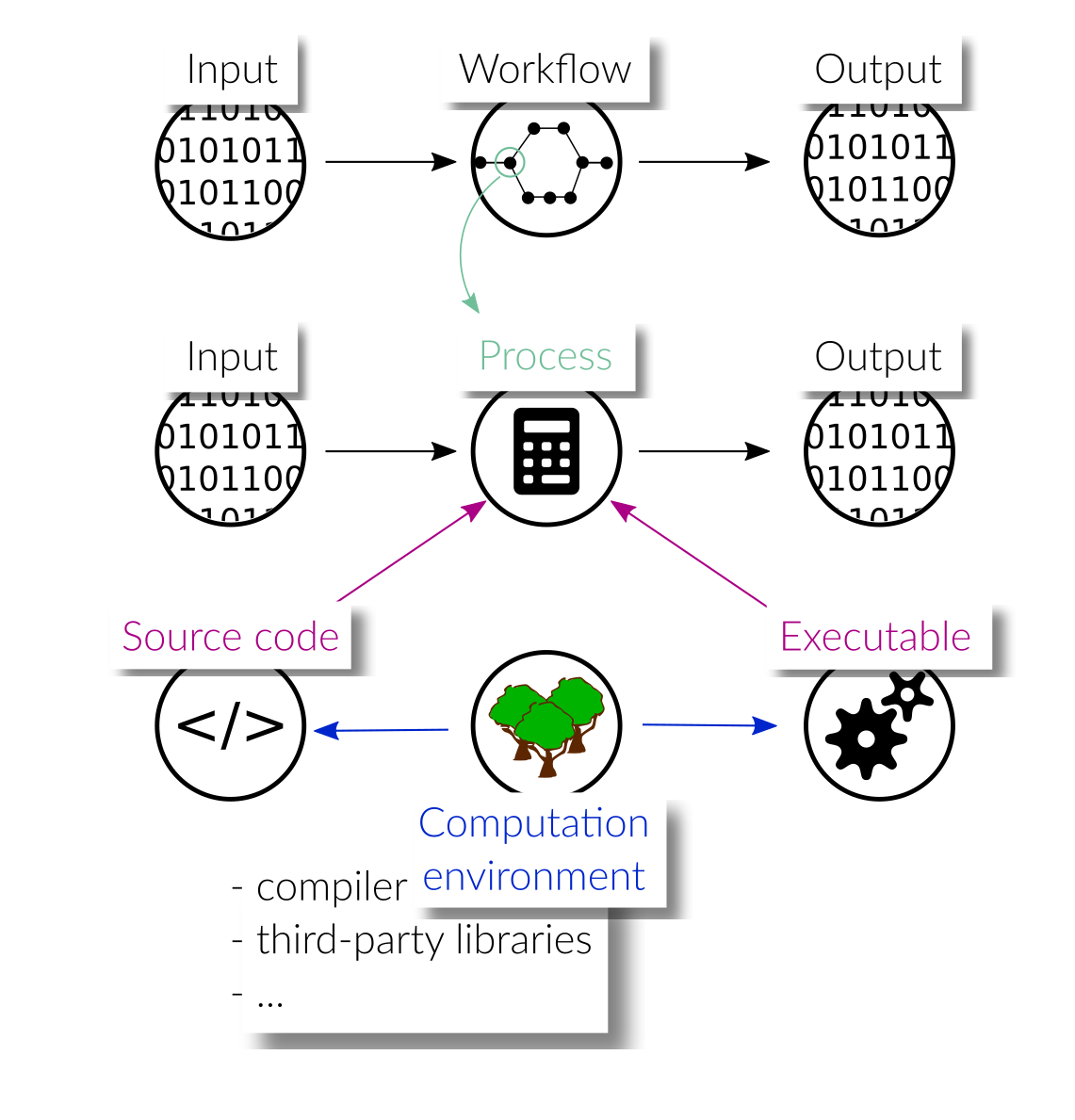}
    \caption{Schematic representation of software-driven scientific workflows.}%
    \label{fig:workflow}
\end{figure}

As shown in~\cref{fig:workflow}, each process in the workflow, just as the workflow itself, takes some input to produce output data.
A more detailed discussion of the different levels of abstractions related to workflows can be found in Griem et al.~\cite{griem2022kadistudio}.
The behavior of a process is primarily determined by the source code that describes it, but may also be influenced by the interpreters/compilers used
for translation or the machines used for execution. Moreover, the source code of a process may carry dependencies to other software packages such that
the behavior of a process possibly depends on their versions.
We use the term \textit{computation environment} to collect all those software dependencies, that is, interpreters and/or compilers as well as third-party libraries and packages that contribute to the computations carried out in a process.
The exact version numbers of all involved packages are crucial, as the workflow may not work with newer or older packages, or, may produce different results.

As outlined in~\cite{Snakemake2021}, WfMSs may be grouped into five classes.
First, tools like \textit{Galaxy}~\cite{galaxy}, \textit{KNIME}~\cite{knime}, and \textit{Pegasus}~\cite{pegasus} provide a graphical user interface~(GUI) to define scientific workflows.
Thus, no programming skills are required and the WfMS is easily accessible for everybody.
With the second group, workflows are defined using a set of classes and functions for generic programming languages (libraries and packages).
This has the advantage that version control (\eg using \textit{Git}~(\url{https://git-scm.com})) can be employed on the workflow.
In addition, the tool can be used without a graphical interface, \eg in a server environment.
Examples of prominent tools are \textit{AiiDA}~\cite{HuberEtal2020,UhrinEtAl2021}, \textit{doit}~\cite{pydoit}, \textit{Balsam}~\cite{balsam_preprint}, \textit{FireWorks}~\cite{fireworks} and \textit{SciPipe}~\cite{scipipe}.
Third, tools like \textit{Nextflow}~\cite{Nextflow2017}, \textit{Snakemake}~\cite{Snakemake2012}, \textit{Bpipe}~\cite{bpipe}, \textit{Guix Workflow Language}~\cite{gwl2022} and \textit{Cluster Flow}~\cite{Ewels2016} express the workflow using a domain specific language~(DSL).
A DSL is a language tailored to a specific problem.
In this context, it offers declarations and statements to implement often occurring constructs in workflow definitions, which
improves the readability and reduces the amount of code.
Moreover, the advantages of the second group also apply for the third group.
In contrast to the definition of the workflow in a programmatic way, the fourth group comprises tools like \textit{Popper}~\cite{popper} and \textit{Argo workflows}~(\url{https://argoproj.github.io/argo-workflows/}) which allow to specify the workflow in a purely declarative way, by using configuration file formats like YAML~\cite{yaml}.
In this case, the workflow specification is concise and can be easily understood, but lacks expressiveness compared to the definition of the workflow using programming languages.
Fifth, there are system-independent workflow specification languages like \textit{CWL}~\cite{cwl_project_june_2022} or \textit{WDL}~(\url{https://github.com/openwdl/wdl}).
These define a declarative language standard for describing workflows, which can then be executed by a number of different engines like \textit{Cromwell}~\cite{cromwell}, \textit{Toil}~\cite{toil}, and \textit{Tibanna}~\cite{tibanna}.

WfMSs can be used to create, execute and monitor workflows.
They can help to achieve reproducibility of research results by avoiding manual steps and automating the execution of the individual processes in the correct order.
More importantly, for a third person to reproduce and reuse the workflow, it needs to be portable, that is, executable on any machine.
Portability can be supported by WfMSs with the integration of package management systems and container technologies, which allow them to automatically
re-instantiate the compute environment.
Another advantage of using WfMSs is the increase in transparency through a clear and readable workflow specification.
Moreover, after completion of the workflow, the tool can help to trace back a computed value to its origin, by logging all inputs, outputs and possibly metadata of all computations.

\section{User stories}%
\label{sec:user_stories}

Starting from user stories that we consider representative for computational science and engineering, a set of requirements is derived that
serves as a basis for the comparison of different WfMSs. In particular, a discussion on how the different tools implement these requirements is provided.

Reproducibility, which is key to transparent research, is the main focus of the first user story (see~\cref{sub:user_story_paper}).
The second user story (see~\cref{sub:user_story_research_group}) deals with research groups that develop workflows in a joint effort where
subgroups or individuals work on different components of the workflow. Finally, the third user story focuses on computational research that involves generating 
and processing large amounts of data, which poses special demands on how the workflow tools organize the data that is created upon workflow execution 
(see~\cref{sub:user_story_complex}).

\subsection{Transparent and reproducible research paper}%
\label{sub:user_story_paper}
\textit{As a researcher, I want to share the code for my paper such that others are able to easily reproduce my results.}

In this user story, the main objective is to guarantee the reproducibility of computational results presented in scientific publications.
Here, reproducibility means that a peer researcher is able to rerun the workflow on some other machine while obtaining
results that are in good agreement with those reported in the publication.
Mere reproduction could also be achieved without a workflow tool, \eg by providing a script that executes the required commands in the right order,
but this comes with a number of issues that may be solved with a standardized workflow description.

First of all, reconstructing the logic behind the generation and processing of results directly from script code is cumbersome and reduces the 
transparency of the research, especially for complex workflows. Second, it is not straightforward for peer researchers to extract certain
processes of a workflow from a script and embed them into a different research project, hence the reusability aspect is poorly addressed with this
solution. Workflow descriptions may provide a remedy to both of these issues, provided that each process in the workflow is defined as a unit with a
clear interface (see~\cref{sub:interfaces}).

While the workflow description helps peers to understand the details behind a research project, it comes with an overhead on the side of the workflow
creator, in particular when using a WfMS for the first time. In the prevalent academic climate, we therefore think that an important aspect of WfMSs is
how easy they are to get started with (see~\cref{sub:ease_of_first_use}).

In the development phase, a workflow is typically run many times until its implementation is satisfactory. With a scripted automation, the entire workflow
is always executed, even if only one process was changed since the last run. Since WfMSs have to know the dependencies between processes, this opens up
the possibility to identify and select only those parts of a workflow that have to be rerun (see~\cref{sub:up_to_dateness}). Besides this, the WfMS can
display to the user which parts are currently being executed, which ones have already been up-to-date, and which ones are still to be picked (see~\cref{sub:monitoring}).

A general issue is that a workflow, or even each process in it, has a specific set of software- and possibly hardware-requirements. This makes both
reproducibility and reusability difficult to achieve, especially over longer time scales, unless the computation environment in which the original study
was carried out is documented in a way that allows for a later re-instantiation. The use of package managers that can export a given environment into
a machine-readable format from which they can then recreate that environment at a later time, may help to overcome this issue. Another promising approach
is to rely on container technologies. WfMSs have the potential to automate the re-instantiation of a computation environment via integration of either
one of the above-mentioned technologies (see~\cref{sub:compute_environment}). This makes it much easier for peers to reproduce and/or reuse parts of a
published workflow.

\subsection{Joint research (group)}%
\label{sub:user_story_research_group}
\textit{As part of a research group, I want to be able to interconnect and reuse components of several different workflows so that everyone may benefit from their colleagues’ work.}

Similar to the previous user story, the output of such a workflow could be a scientific paper. However, this user story explicitly considers interdisciplinary
workflows in which the reusability of individual components/modules is essential. Each process in the workflow may require a different expertise and hence modularity and a 
common framework is necessary for an efficient collaboration.

Many of the difficulties discussed in the previous user story are shared in a joint research project. However, the collaborative effort in which the
workflow description and those of its components are developed promotes the importance of clear interfaces (see~\cref{sub:interfaces}) to ease communication and an intuitive
dependency handling mechanism (see~\cref{sub:compute_environment}).

Another challenge here is that such workflows often consist of heterogeneous models of different complexity, such as large computations requiring 
high-performance computing~(HPC), preprocessing of experimental data or postprocessing analyses. 
Due to this heterogeneity, it may be beneficial to outsource computationally demanding tasks to HPC systems, while executing cheaper tasks locally 
(see~\cref{sub:execution}). Workflows with such computationally expensive tasks can also strongly benefit from effective caching mechanisms and the
reuse of cached results wherever possible (see~\cref{sub:up_to_dateness}).

Finally, support for a hierarchical embedding of sub-workflows (possibly published and versioned) in another workflow is of great benefit as this allows for an 
easy integration of improvements made in the sub-workflows by other developers (see~\cref{sub:hierarchical_composition_of_workflows}).

\subsection{Complex hierarchical computations}%
\label{sub:user_story_complex}
\textit{As a materials scientist, I want to be able to automate and manage complex workflows so I can keep track of all associated data.}

Workflows in which screening or parameter sweeps are required typically involve running a large number of simulations. Moreover, these
workflows are often very complex with many levels of dependencies between the individual tasks. Good data management that provides access to the
full provenance graph of all data can help to retain an overview over the large amounts of data
produced by such workflows (see~\cref{sub:data_provenance}). For instance, the data management could be such that desired information may be efficiently
extracted via query mechanisms.

Due to the large amount of computationally demanding tasks in such workflows, it is helpful if some computations can be outsourced to HPC systems
(see~\cref{sub:execution}) with a clean way of querying the current status during the typically long execution times (see~\cref{sub:monitoring}).

\section{Specific requirements on workflow management systems}%
\label{sec:requirements}
The user stories described above allow us to identify 11 requirements on WfMSs. 
They will be described in the following and serve as evaluation criteria for the individual WfMSs discussed in~\cref{sec:tool_comparison}.

\subsection{Support for job scheduling system}%
\label{sub:execution}
As already mentioned, the main task of a WfMS is to automatically execute the processes of a workflow in the correct order such that the dependencies between them are satisfied.
However, processes that do not depend on each other may be executed in parallel in order to speed up the overall computation time.
This requirement focuses on the ability of a workflow tool to distribute the computations on available resources.
Job scheduling systems like \eg Slurm (also commonly referred to as batch scheduling or batch systems) are often used to manage computations to be run and their resource
requirements (number of nodes, CPUs, memory, runtime, etc.).
Therefore, it is of great benefit if WfMSs support the integration of widely-used batch systems such that users can specify and also observe
the used resources alongside other computations that were submitted to their batch system in use.
Besides this, this requirement captures the ability of a WfMS to outsource computations to a remote machine, \eg a HPC cluster or cloud.
For traditional HPC cluster systems it is usually necessary to transfer input and output data between the local system and the cluster system.
This can be done using the secure shell protocol (SSH) and a WfMS may provide the automated transfer of a job's associated data.
Ideally, the workflow can be executed anywhere without changing the workflow definition itself, but only the runtime arguments or a configuration file.
The fulfillment of this requirement is evaluated by the following criteria:
\begin{itemize}
    \item[\ci{1}{3}] The workflow system supports the execution of the workflow on the local system.
    \item[\ci{2}{3}] The workflow system supports the execution of the workflow on the local system via a batch system.
    \item[\ci{3}{3}] The workflow system supports the execution of the workflow via a batch system on the local or a remote system.
\end{itemize}

\subsection{Monitoring}
\label{sub:monitoring}
Depending on the application, the execution of scientific workflows can be very time-consuming. This can be caused by compute-intensive processes
such as numerical simulations, or by a large number of short processes that are executed many times. In both cases, it can be very helpful to
be able to query the state of the execution, that is, which processes have been finished, which processes are currently being executed, and which
are still pending. A trivial way of such monitoring would be, for instance, when the workflow is started in a terminal which is kept open to inspect
the output written by the workflow system and the running processes. However, ideally, the workflow system allows for submission of the workflow in
the form of a process running in the background, while still providing means to monitor the state of the execution. 
For this requirement, two criteria are distinguished:
\begin{itemize}
    \item[\ci{1}{2}] The only way to monitor the workflow is to watch the console output.
    \item[\ci{2}{2}] The workflow system provides a way to query the execution status at any time.
\end{itemize}

\subsection{Graphical user interface}%
\label{sub:graphical_user_interface}

Independent of a particular execution of the workflow, the workflow system may provide facilities to visualize the graph of the workflow, indicating the
mutual dependencies of the individual processes and the direction of the flow of data. One can think of this graph as the template for the data provenance
graph. This visualization can help in conveying the logic behind a particular workflow, making it easier for other researchers to understand and possibly
incorporate it into their own research. The latter requires that the workflow system is able to handle hierarchical workflows, that is, 
workflows that contain one or more sub-workflows as processes (see~\cref{sub:hierarchical_composition_of_workflows}).
Beyond a mere visualization, a GUI may allow for visually connecting different
workflows into a new one by means of drag \& drop.
We evaluate the features of a graphical user interface by means of the following three criteria:
\begin{itemize}
    \item[\ci{1}{3}] The workflow system provides no means to visualize the workflow
    \item[\ci{2}{3}] The workflow system or third-party tools allow to visualize the workflow definition
    \item[\ci{3}{3}] The workflow system or third-party tools provide a GUI that enables users to graphically create workflows
\end{itemize}

\subsection{Data provenance}
\label{sub:data_provenance}
The data provenance graph contains, for a particular execution of the workflow, which data and processes participated in the generation of a particular
piece of data. Thus, this is closely related to the workflow itself, which can be thought of as a template for how that data generation should take place.
However, a concrete realization of the workflow must contain information on the exact input data, parameters and intermediate results,
possibly along with meta information on the person that executed the workflow, the involved software, the compute resources used and the time 
it took to finish. Collection of all relevant information, its storage in machine-readable formats and subsequent publication alongside the data can be very 
useful for future researchers in order to understand how exactly the data was produced. 
Ideally, the workflow system has the means to automatically collect 
this information upon workflow execution, which we evaluate using the following criteria:
\begin{itemize}
    \item[\ci{1}{2}] The workflow system provides no means to export relevant information from a particular execution
    \item[\ci{2}{2}] The workflow system stores all results (also intermediate) together with provenance metadata about how they were produced
\end{itemize}

\subsection{Compute environment}%
\label{sub:compute_environment}
In order to guarantee interoperability and reproducibility of scientific workflows, the workflows need to be executable by others.
Here, the re-instantiation of the compute environment (installation of libraries or source code) poses the main challenge.
Therefore, it is of great use if the workflow tool is able to automatically deploy the software stack (on a per workflow or per process basis) by means of a package manager (\eg conda \url{https://conda.io/}) or that running processes in a container (\eg Docker \url{https://www.docker.com}, Apptainer \url{https://apptainer.org} (formerly Singularity)) is integrated in the tool.
The automatic deployment of the software stack facilitates the execution of the workflow.
However, it does not (always) enable reusage, that is, the associated software can be understood, modified, built upon or incorporated into other software~\cite{fair4rs}.
For instance, if a container image is used, it is important that the container build recipe (\eg Dockerfile) is provided.
This increases the reusability as it documents how a productive environment, suitable to execute the given workflow or process, can be set up.
The author of the workflow, however, is deemed to be responsible for the documentation of the compute environment. 
For this requirement, we define the following evaluation criteria:
\begin{itemize}
    \item[{\ci{1}{3}}] The automatic instantiation of the compute environment is not intended.
    \item[\ci{2}{3}] The workflow system allows the automatic instantiation of the compute environment on a per workflow basis.
    \item[\ci{3}{3}] The workflow system allows the automatic instantiation of the compute environment on a per process basis.
\end{itemize}

\subsection{Hierarchical composition of workflows}%
\label{sub:hierarchical_composition_of_workflows}
A workflow consists of a mapping between a set of inputs (could be empty) and a set of outputs, whereas 
in between a number of processes are performed. 
Connecting the output of one workflow to the input of another workflow results in a new, longer workflow.
This is particularly relevant in situations, where multiple people share a common set of procedures (\eg common pre- and postprocessing routines).
In this case, copying the preprocessing workflow into another one is certainly always possible, but does not allow to jointly perform 
modifications and work with different versions. Moreover, a composition might also require to define separate compute environments for each sub-workflow (e.g. 
using docker/singularity or conda). Executing all sub-workflows in the same environment might not be possible because each sub-workflow might use different tools or 
even the same tools but with different versions (\eg python2 vs. python3).
This promotes the importance of supporting heterogeneous compute environments, which is reflected in the evaluation criteria for this requirement:
\begin{itemize}
    \item[\ci{1}{3}] The workflow system does not allow the composition of workflows.
    \item[\ci{2}{3}] The workflow system allows to embed a workflow into another one for a single compute environment (homogeneous composition).
    \item[\ci{3}{3}] The workflow system allows to embed a workflow into another one for arbitrary many (on a per process basis) compute environments (hierarchical composition).
\end{itemize}

\subsection{Interfaces}
\label{sub:interfaces}
In a traditional file-based pipeline, the output files produced by one process are used as inputs to a subsequent process.
However, it is often more convenient to pass non-file output (\eg float or integer values) directly from one process to another without the creation of intermediate files.
In this case, it is desirable that the workflow tool is able to check the validity of the data (\eg the correct data type) to be processed.
Furthermore, this defines the interface for a process more clearly and makes it easier for someone else to understand how to use, adapt or extend the workflow/process.
In contrast, in a file-based pipeline, this is usually not the case since a dependency in form of a file does not give information about the type of data 
contained in that file.
We distinguish these different types of interfaces by the following criteria:
\begin{itemize}
    \item[\ci{1}{3}] The workflow system is purely file-based and does not define interface formats.
    \item[\ci{2}{3}] The workflow system allows for passing file and non-file arguments between processes.
    \item[\ci{3}{3}] The workflow system allows for defining strongly-typed process interfaces, supporting both file and non-file arguments.
\end{itemize}

\subsection{Up-to-dateness}
\label{sub:up_to_dateness}
There are different areas for the application of workflows. On the one hand, people might use a workflow to define a single piece of reproducible code that, when executed, always returns the same result. Based on that, they might start a large quantity of different jobs and use the workflow system to perform this task. Another area of application is the constant development within the workflow (e.g. exchanging processes, varying parameters or even modifying the source code of a process) until a satisfactory result is obtained. The two scenarios require a slightly different behavior of the workflow system. In the first scenario, all runs should be kept in the data provenance graph with a documentation of how each result instance has been obtained (e.g. by always documenting the codes, parameters, and processes). If identical runs (identical inputs and processes should result in the same output) are detected, a recomputation should be avoided and the original output should be linked in the data provenance graph. The benefit of this behavior certainly depends on the ratio between the computation time for a single process compared to the overhead to query the data base.

However, when changing the processes (e.g. coding a new time integration scheme or a new constitutive model), the workflow system should rather behave like a build system (such as make) - only recomputing the steps that are changed or that depend on these changes. In particular for complex problems, this allows to work with complex dependencies without manually triggering computations and results in automatically recomputing only the relevant parts. An example is a paper with multiple figures where each is a result of complex simulations that in itself depend on a set of general modules developed in the paper. The ``erroneous'' runs are usually not interesting and should be overwritten.

How this is handled varies between the tools, yielding the following evaluation criteria:
\begin{itemize}
    \item[\textbf{R}] The complete workflow is always \textbf{R}ecomputed.
    \item[\textbf{L}] A new entry in the data provenance graph is created which \textbf{L}inks the previous result (without the need to recompute already existing results).
    \item[\textbf{U}] Only the parts are recreated (\textbf{U}pdated) that are not up-to-date. This usually reduces the overhead to store multiple instances of the workflow, but at the same time also prevents - without additional effort (e.g. when executing in different folders) computing multiple instances of the same workflow.
\end{itemize}

\subsection{Ease of first use}%
\label{sub:ease_of_first_use}
Although this is not a requirement per-se, it is beneficial if the workflow system has an intuitive syntax/interface and little work is required for a new user 
to define a first workflow. Research applications typically have a high intrinsic complexity, and therefore, the complexity added by the workflow management 
should be as small as possible. Evaluation criteria:
\begin{itemize}
    \item[\ci{1}{3}] difficult
    \item[\ci{2}{3}] intermediate
    \item[\ci{3}{3}] easy
\end{itemize}

\subsection{Manually editable workflow definition}%
\label{sub:manually_editable_workflow_definition}
While it can be beneficial to create and edit workflows using a GUI (see~\cref{sub:graphical_user_interface}), it may be important that 
the resulting workflow description is given in a human-readable format. This does not solely mean that the definition should be a text file, but also
that the structure (\eg indentation) and the naming are comprehensive. This facilitates version-controlling with git, in particular the code review
process. Moreover, this does not force all users and/or developers to rely on the GUI.
Evaluation criteria:
\begin{itemize}
    \item[\ci{1}{3}] The workflow description is a binary file.
    \item[\ci{2}{3}] The workflow description is a text file but hard to interpret by humans.
    \item[\ci{3}{3}] The workflow description is a fully human-readable file format.
\end{itemize}

\subsection{Platform for publishing and sharing workflows}%
\label{sub:platform_for_publishing_and_sharing_workflows}
The benefit of a workflow system is already significant when using it for individual research such as the development of an individual's paper or reproducing 
the paper someone else has written, when their data processing pipeline is fully reproducible, documented and published. However, the benefit can be even more 
increased if people are able to jointly work on (sub-)workflows together; particularly when a hierarchical workflow system is used. Even though workflows can
easily be shared together with the work (e.g. in a repository), it might be beneficial to provide a platform that allows to publish documented workflows with a 
search and versioning functionality. This feature is not part of the requirement matrix to compare the different tools, but we consider a documentation of 
these platforms in the subsequent section as a good starting point for further research (exchange).

\section{Simple use case}%
\label{sec:simple_use_case}

\begin{figure}[htpb]
	\centering
	\includegraphics[width=.8\textwidth]{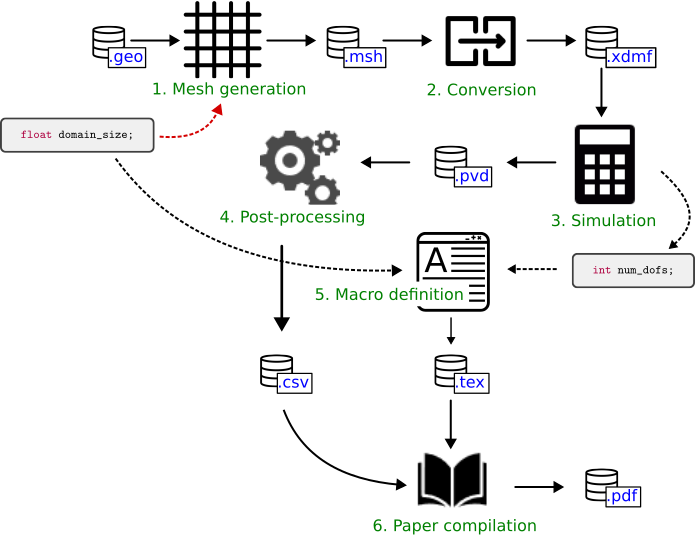}
	\caption{Task dependency graph of the simple use case. Mapping of input and output data is indicated with black arrows with solid lines. A dashed line refers to non-file input or output (parameters). Here, red color is used to distinguish user input from automatic data transfer.}%
	\label{fig:dag_simple_use_case}
\end{figure}

A simple exemplary use case was defined in order to analyze and evaluate the
different workflow tools with respect to the requirements stated in~\cref{sec:requirements}.
This example is considered to be representative for many problems simulating
physical processes in engineering science using numerical discretization techniques.
It consists of six steps, as shown in~\cref{fig:dag_simple_use_case}:

\begin{enumerate}
  \item generation of a computational mesh (Gmsh)
  \item mesh format conversion (MeshIO)
  \item numerical simulation (FEniCS)
  \item post-processing of the simulation results (ParaView)
  \item preparation of macro definitions (Python)
  \item compilation of a paper into a \textit{.pdf}  file using the simulation results (Tectonic)
\end{enumerate}

The workflow starts from a given geometry on which the simulation should be carried out and generates a computational mesh in the first step using Gmsh~\cite{GeuzaineRemacle2009}.
Here, the user can specify the size of the computational domain by a float value \texttt{domain\_size}.
The resulting mesh file format is not supported by FEniCS~\cite{AlnaesEtal2015}, which is the software that we are using for the simulation carried out in the third step.
Therefore, we convert the mesh file in the second step of the workflow from \textit{.msh} to \textit{.xdmf} using the python package MeshIO~\cite{meshio534}.
The simulation step yields result files in VTK file format~\cite{Schroeder2006} and returns the number of degrees of freedom used by the simulation as an integer value \texttt{num\_dofs}.
The VTK files are further processed using the python application programming interface~(API) of ParaView~\cite{Ahrens2005ParaViewAE}, which yields the data of a plot-over-line of the numerical solution across the domain in \textit{.csv} file format.
This data, together with the values for the domain size and the number of degrees of freedom, is inserted into the paper and compiled into a \textit{.pdf} file using the \LaTeX\ engine Tectonic~\cite{tectonic} in the final step of the workflow.

Most steps transfer data among each other via files, but we intentionally built in the transfer of the number of degrees of freedom as an integer value to check how well such a situation can be handled by the tools.
Example implementations of the simple use case for various tools are available in a public repository~\cite{NFDI4Ing_Scientific_Workflow_2022}.

\section{Tool comparison}%
\label{sec:tool_comparison}
In this section, the selected WfMSs and their most important features are described and set in relation to the requirements defined in~\cref{sec:requirements}.
We note that to the best of our knowledge, existing add-on packages to the individual WfMSs are as well considered.

\subsection{AiiDA}%
\label{sub:aiida}
\textit{AiiDA}~\cite{HuberEtal2020,UhrinEtAl2021}, the automated interactive infrastructure and database for computational science, is an open source Python infrastructure.
With \textit{AiiDA}, workflows are written in the Python programming language and managed and executed using the associated command line interface \enquote{\textit{verdi}}.

\textit{AiiDA} was designed for use cases that are more focused on running heavy simulation codes on heterogeneous compute hardware.
Therefore, one of the key features of \textit{AiiDA} is the HPC interface.
It supports the execution of (sub-) workflows on any machine and most resource managers are integrated.
In case of remote computing resources, any data transfer, retrieval and storing of the results in the database or status checking is handled by the \textit{AiiDA} daemon process.
Another key feature is \textit{AiiDA}'s workflow writing system which provides strongly typed interfaces and allows for easy composition and reuse of workflows.
Moreover, \textit{AiiDA} automatically keeps track of all inputs, outputs and metadata of all calculations, which can be exported in the form of provenance graphs.

\textit{AiiDA}'s workflow system enables to easily compose workflows, but \textit{AiiDA} lacks in providing the compute environment, such that the composition of heterogeneous 
workflows is challenging since it requires the installation of software dependencies of the workflow on any machine that should be used with \textit{AiiDA}\@. 
The reason for this may be the challenges in using conda or containers on HPC systems.
On traditional HPC systems the preferred way of running software is to use the provided module system to compile specific application code.
The system may be isolated, such that missing access to the internet prevents installing conda environments or downloading container images.
Moreover, successfully using container technology as an MPI-distributed application across several nodes seems to be a technical challenge due to compatibility issues in the MPI configuration and certain Infiniband drivers.

In addition to that, running external codes with \textit{AiiDA} requires the implementation of an \textit{AiiDA} plugin which instructs \textit{AiiDA} on how to run that code.
This poses an additional overhead if the application code changes frequently during development of the workflow.
Also, in the special case of FEniCS (see~\cref{sec:simple_use_case}), which can be used to solve partial differential equations and therefore covers a wide spectrum of applications, it is very difficult to define a general plugin interface which covers all models.
We note that due to this use case which is rather different from the use cases that \textit{AiiDA} was designed for, the implementation of the simple use case 
(see~\cref{sec:simple_use_case}) uses \enquote{aiida-shell}~\cite{Huber_aiida-shell_2022}, an extension to the \textit{AiiDA} core package which makes running shell 
commands easy.
While this is convenient to get a workflow running quickly, this leads to an undefined process interface since this was the purpose of the aforementioned plugin for an external code.
Considering the points above, compared to the other tools, the learning curve with \textit{AiiDA} is fairly steep.
In contrast to file-based workflow management systems, \textit{AiiDA} defines data types for any data that should be stored in the database.
Consequently, non-file based inputs are well defined, but this is not necessarily the case for file data.

In terms of the requirements defined in~\cref{sec:requirements}, \textit{AiiDA}'s strong points are execution, monitoring and provenance.
Due to the possibility to export provenance graphs, also level two of the requirement \enquote{Graphical user interface} is reached.
Lastly, caching can be enabled in \textit{AiiDA} to save computation time.
Caching in \textit{AiiDA} means, that the database will be searched for a calculation of the same hash and if this is the case, the same outputs are reused.

\subsection{Common Workflow Language}%
\label{sub:cwl}

\enquote{\textit{Common Workflow Language} (\textit{CWL})~\cite{Amstutz2016} is an open standard for describing how to run command line tools
and connect them to create workflows}~(\url{https://www.commonwl.org/}).
One benefit of it being a standard is that workflows expressed in \textit{CWL} do not have to be executed by a particular
workflow engine, but can be run by any engine that is able to parse the standard. In fact, there exist a number of
workflow engines that support \textit{CWL} workflows, \eg the reference implementation \textit{cwltool}~(\url{https://github.com/common-workflow-language/cwltool}), \textit{Toil}~\cite{toil}
or \textit{StreamFlow}~\cite{Colonnelli_StreamFlow_cross-breeding_cloud_2021}.

\textit{CWL} was designed with a focus on data analysis using command line programs. To create a workflow, each of the
command line programs is \enquote{wrapped} in a \textit{CWL} description, defining what inputs are needed, what outputs are
produced and how to call the underlying program. Typically, this step also reduces the possibly large number
of options of the underlying command line tool to a few options or inputs that are relevant for the particular
task of the workflow. In a workflow, the wrapped command line 
tools can be defined as individual processes, and the
outputs of one process can be mapped to the inputs of other processes. This information is enough for the interpreter
to build up the dependency graph, and processes that do not depend on each other may be executed in parallel. A process
can also be another workflow, thus, hierarchical workflow composition is possible. Moreover, there exist workflow
engines for \textit{CWL} that support using job managers like \eg Slurm~\cite{slurm_2003}.

The \textit{CWL} standard also provides means to specify the software requirements of a process. For instance, one
can provide the URL of a docker image or docker file to be used for the execution of a process. In case of the latter,
the image is automatically built from the provided docker file, which itself contains the information on all required
software dependencies. Besides this, the \textit{CWL} standard contains language features that allow listing software
dependencies directly in the description of a workflow or process, and workflow engines may automatically make
these software packages available upon execution. As one example, the current release of \textit{cwltool}
supports the definition of software requirements in the form of \eg \textit{Conda}
packages that are then automatically installed when the workflow is run (see \eg our implementation and the respective
pipelines at~\cite{NFDI4Ing_Scientific_Workflow_2022}).

In contrast to workflow engines that operate within a particular programming language, the transfer of data from one process to
another cannot occur directly via memory with \textit{CWL}. For instance, if the result of a process is an integer value, this value
has to be read from a file produced by the process, or, from its console output. However, this does not have to be done in a
separate process or by again wrapping the command line tool inside some script, since \textit{CWL} supports the definition of
inline JavaScript code that is executed by the interpreter. This allows retrieving, for instance, integer or floating point
return values from a process with a small piece of code.

\textit{CWL} requires the types of all inputs and outputs to be specified, which has the benefit that the interpreter can do type checks
before the execution of the workflow. A variety of primitive types, as well as arrays, files or directories, are available.
Files can refer to local as well as online resources, and in the case of the latter, resources are automatically fetched
and used upon workflow execution.

There exist a variety of tools built around the \textit{CWL} standard, such as the Rabix Composer~(\url{https://rabix.io/}) for visualizing and
composing workflows in a GUI. Besides that and as mentioned before, there are several workflow engines
that support \textit{CWL} and some of which provide extra features. For instance, \textit{cwltool} allows for tracking provenance information
of individual workflow runs. However, to the best of our knowledge, there exists no tool that automatically checks which
results are up-to-date and do not have to be reproduced (see~\cref{sub:up_to_dateness}).

The \textit{CWL} standard allows to specify the \textit{format} of an input or output file by means of an \textit{IRI} 
(Internationalized Resource Identifier) that points to online-available resource where the file format is defined.
For processes whose output files are passed to the inputs of subsequent jobs, the workflow engine can use this information
to check if the formats match. To the best of our knowledge, \textit{cwltool} does so by verifying that the \textit{IRI}s
are identical, or performs further reasoning in case the \textit{IRI}s point to classes in ontologies (see, for instance, the
class for the JSON file format in the EDAM ontology at \href{http://edamontology.org/format_3464}{edamontology.org/format\_3464}). Such reasoning can make
use of defined relationships between classes of the ontology to determine file format compatibility.
For more information on file format specifications in \textit{CWL} see \href{https://www.commonwl.org/user_guide/topics/file-formats.html}{commonwl.org/user\_guide/topics/file-formats.html}.

\subsection{doit}%
\label{sub:doit}

\enquote{\textit{doit} comes from the idea of bringing the power of build-tools to execute any kind of task}~\cite{pydoit}.
The automation tool \textit{doit} is written in the Python programming language.
In contrast to systems which offer a GUI, knowledge of the programming language is required.
However, it is not required to learn an additional API since task metadata is returned as a Python dictionary.
Therefore, we consider this as very easy to get started quickly.

With \textit{doit}, any shell command available on the system or python code can be executed.
This also includes the execution of processes on a remote machine, although all necessary steps (\eg connecting to the remote via SSH) need to be defined by the user.
In general, such behavior as described in~\cref{sub:execution} is possible, but it is not a built-in feature of \textit{doit}.
Also, \textit{doit} does not intend to provide the compute environment.
Therefore, while in general the composition of workflows (see~\cref{sub:hierarchical_composition_of_workflows}) is easily possible via python imports, this only works for a single environment.
The status of the execution can be monitored via the console.
Here, \textit{doit} will skip the execution of processes which are up-to-date and would produce the same result of a previous execution.
To determine the correct order in which processes should be executed, \textit{doit} also creates a directed acyclic graph~(DAG) which could be used to visualize dependencies between processes using \enquote{\textit{doit-graph}}~(\url{https://github.com/pydoit/doit-graph}), an extension to \textit{doit}.
For each run (specific instance of the workflow), \textit{doit} will save the results of each process in a database.
However, the tool does not provide control over what is stored in the database.
On the one hand, \textit{doit} allows to pass results of one process as input to another process directly, without creating intermediate files, so it is not purely file-based.
On the other hand, the interface for non-file based inputs does not define the data type.

\subsection{Guix Workflow Language}
The \textit{Guix Workflow Language} (\textit{GWL})~\cite{gwl2022} is an extension to the open
source package manager GNU Guix~\cite{Courtes2013}. \textit{GWL} leverages several
features from Guix, chief among these is the compute environment management.
Like Guix, \textit{GWL} only supports GNU/Linux systems.

\textit{GWL} can automatically construct an execution graph from the workflow
process input/output dependencies but also allows a manual specification.
Support for HPC schedulers via DRMAA\footnote{Distributed Resource Management
Application API \url{https://www.drmaa.org}} is also available.

\textit{GWL} doesn't provide a graphical user interface, interactions are carried out
using a command-line interface in a text terminal. Monitoring is also only
available in the form of simple terminal output.

There is support to generate a GraphViz (see \eg~\url{https://graphviz.org})
description of the workflow, which allows basic visualization of a workflow.
Although not conveniently exposed\footnote{\textit{GWL} doesn't provide a command to
export provenance graphs in any way, instead Guix needs to be queried for build
instruction, dependency graphs and similar provenance information of a workflows
software packages}, \textit{GWL} has a noteworthy unique feature inherited from Guix:
precise software provenance tracking. Guix contains complete build instructions
for every package (including their history through git), which enables
accounting of source code and the build process, like for example compile
options, of all tools used in the workflow. Integrity of this information is
ensured through cryptographic hash functions. This information can be used to
construct data provenance graphs with a high level of integrity (basically all
userspace code of the compute environment can be accounted
for~\cite{Courtes2022}).

\textit{GWL} uses Guix to setup compute environments for workflow processes.
Each process is executed in an isolated\footnote{By default, lightweight
isolation is setup by limiting the \texttt{PATH} environment variable to the
compute environment. Stronger isolation via Linux containers is also optionally
available.} compute environment in which only specified software
packages are available. This approach minimizes (accidential)
side-effects from system software packages and improves workflow
reproducibility. Interoperability also benefits from this approach, since a Guix
installation is the only requirement to execute a workflow on another machine.
As Guix provides build instructions for all software packages, it should be
easily possible to recreate compute environments in the future, even if the
originally compiled binaries have been deprecated in the meanwhile
(see~\cite{akhlagi2022} for a discussion about long-term reproducibility).

Composition of workflows is possible, workflows can be imported into other
workflows. Composition happens either by extracting individual processes
(repurposing them in a new workflow) or by appending new processes onto the
existing workflow processes.

\textit{GWL} relies exclusively on files as interface to workflow processes. There's no
support to exchange data on other channels, as workflow processes are executed
in isolated environments.

Like other workflow tools, \textit{GWL} caches the result of a workflow process
using the hash of its input data. If a cached result for the input hash value
exists, the workflow processes execution is skipped.

\textit{GWL} is written in the Scheme~\cite{Sperber2009} implementation GNU
Guile~\cite{guile2022}, but in addition to Scheme, workflows can also be defined
in wisp~\cite{wisp2015}, a variant of Scheme with significant whitespace. wisp
syntax thus resembles Python, which is expected to flatten the learning curve a
bit for scientific audience. However, error messages are very hard to read
without any background in Scheme. On first use, \textit{GWL} will be very difficult in
general. This problem is acknowledged by the \textit{GWL} authors and might be subject to
improvements in the future.

As both wisp and Scheme code is almost free of syntactic noise in general,
workflows are almost self-describing and easily human-readable.

In summary, \textit{GWL} provides a very interesting and sound set of features especially
for reproducibility and interoperability. These features come at the cost of a
Guix installation, which requires administrator privileges. The workflow
language is concise and expressive, but error messages are hard to read. At the
current stage, \textit{GWL} can only be recommended to experienced scheme programmers or
to specialists with high requirements on software reproducibility and integrity.

\subsection{Nextflow and Snakemake}%
\label{sub:nextflow_and_snakemake}%

With \textit{Nextflow}~\cite{Nextflow2017} and \textit{Snakemake}~\cite{Snakemake2021}, the workflow is defined using a DSL which is an extension to a generic programming language (Groovy for \textit{Nextflow} and Python for \textit{Snakemake}).
Moreover, \textit{Nextflow} and \textit{Snakemake} also allow to use the underlying programming language to generate metadata programmatically.
Thus, authoring scientific workflows with \textit{Nextflow} or \textit{Snakemake} is very easy.

The process to be executed is usually a shell command or an external script.
The integration with various scripting languages is an import feature of \textit{Snakemake} as well as \textit{Nextflow}, which encourages readable modular code for downstream plotting and summary tasks.
Also boilerplate code for command line interfaces~(CLIs) in external scripts can be avoided.
Another feature of \textit{Snakemake} is the integration of Jupyter notebooks, which can be used to interactively develop components of the workflow.

Both tools implement a CLI to manage and run workflows.
By default, the status of the execution is monitored via the console.
With \textit{Nextflow}, it is possible to monitor the status of the execution via a weblog.
\textit{Snakemake} supports an external server to monitor the progress of submitted workflows.

With regard to the execution of the workflow~(\cref{sub:execution}), the user can easily run the workflow on the local machine and the submission via a resource manager~(\eg Slurm, Torque) is integrated.
Therefore, individual process resources can be easily defined with these tools if the workflow is submitted on a system where a resource manager is installed, \ie on a traditional HPC cluster system.
Despite this, only level two of the defined criteria is met for \textit{Nextflow}, since the execution of the workflow on a remote machine and the accompanied transfer of data is not handled by the tool.
For \textit{Snakemake}, if the CLI option \enquote{default-remote-provider} is used, all input and output files are automatically down- and uploaded to the defined remote storage, such that no workflow modification is necessary.

The requirement \enquote{up-to-dateness} is handled differently by \textit{Nextflow} and \textit{Snakemake}.
By default, \textit{Nextflow} recomputes the complete workflow, but with a single command-line
option existing results are retrieved from the cache and linked such that a re-execution is not
required.
In this case, \textit{Nextflow} allows storing multiple instances of the same workflow upon variation of a configuration parameter.
\textit{Snakemake} will behave like a build tool in this context and skip the re-execution of processes whose targets already exist and update any process whose 
dependencies have changed.

A strong point of \textit{Nextflow} and \textit{Snakemake} is the integration of the conda package management system and container technologies like docker.
For example, the compute environment can be defined for each process based on a conda environment specification file or a certain docker image.
Upon execution of the workflow, the specified compute environment is re-instantiated automatically by the workflow tool, making it very easy
to reproduce results of or built upon existing workflows.
Furthermore, since the tool is able to deploy the software stack on a per process basis, the composition of hierarchical workflows as outlined in~\cref{sub:hierarchical_composition_of_workflows} is possible.

Similar to \textit{doit}, both tools do not provide a GUI to graphically create and modify workflows.
However, a visualization of the workflow, \ie a dependency graph of the processes, can be exported.
Moreover, it is possible to export extensive reports detailing the provenance of the generated data.

\textit{Nextflow} and \textit{Snakemake} can also be regarded as file-based workflow management systems.
Therefore, interface formats, \ie class structures or types of the parameters passed from one process to the subsequent one, are not clearly defined.

\subsection{Evaluation matrix}%
\label{sub:summary_of_the_tools}

The evaluation of the workflow tools provided in~\cref{sec:tool_comparison} in terms of the requirements described in~\cref{sec:requirements} on 
the example of the workflow outlined in~\cref{sec:simple_use_case} yields the evaluation matrix depicted in~\cref{tab:eval_tools}. 
\begin{table}[htb]

\caption{Evaluation of the workflow tools.}%
\label{tab:eval_tools}
\begin{tabularx}{\textwidth}{lXXXXXX}\toprule
	Requirement              & \multicolumn{6}{c}{Workflow tool} \\\cmidrule(lr){2-7}
                             & \textit{AiiDA}     & \textit{CWL}       & \textit{doit}      & \textit{GWL}       & \textit{Nextflow}  & \textit{Snakemake} \\ \midrule
       Job scheduling system & \ci{3}{3} & \ci{2}{3} & \ci{1}{3} & \ci{2}{3} & \ci{2}{3} & \ci{3}{3} \\
	              Monitoring & \ci{2}{2} & \ci{2}{2} & \ci{1}{2} & \ci{1}{2} & \ci{1}{2} & \ci{1}{2} \\
	Graphical user interface & \ci{2}{3} & \ci{3}{3} & \ci{2}{3} & \ci{1}{3} & \ci{2}{3} & \ci{2}{3} \\
	              Provenance & \ci{3}{3} & \ci{2}{3} & \ci{1}{3} & \ci{1}{3} & \ci{2}{3} & \ci{2}{3} \\
	     Compute environment & \ci{1}{3} & \ci{3}{3} & \ci{1}{3} & \ci{3}{3} & \ci{3}{3} & \ci{3}{3} \\
                 Composition & \ci{2}{3} & \ci{3}{3} & \ci{2}{3} & \ci{3}{3} & \ci{3}{3} & \ci{3}{3} \\
	      Process interfaces & \ci{2}{3} & \ci{3}{3} & \ci{1}{3} & \ci{1}{3} & \ci{1}{3} & \ci{1}{3} \\
	          Up-to-dateness & L         & R         & U         & U         & L         & U         \\
	       Ease-of-first-use & \ci{1}{3} & \ci{2}{3} & \ci{3}{3} & \ci{1}{3} & \ci{3}{3} & \ci{3}{3} \\
	       Manually editable & \ci{3}{3} & \ci{3}{3} & \ci{3}{3} & \ci{3}{3} & \ci{3}{3} & \ci{3}{3} \\
	\bottomrule
\end{tabularx}
\end{table}

\section{Summary}

In this work, six different WfMSs (\textit{AiiDA}, \textit{CWL}, \textit{doit}, \textit{GWL}, \textit{Nextflow} and \textit{Snakemake}) are studied. 
Their performance is evaluated based on a set of requirements derived from three typical user stories in the field of computational science and engineering.
On the one hand, the user stories are focusing on facilitating the development process, and on the other hand on the possibility of reusing and reproducing results obtained using research software.
The choice for one WfMS or the other is strongly subjective and depends on the particular application and the preferences of its developers.
The overview given in~\cref{tab:eval_tools} together with the assessments in~\cref{sec:tool_comparison} may only serve as a basis for an individual decision making.

For researchers that want to start using a WfMS, an important factor is how easy it is to get a first workflow running.
For projects that are written in Python, a
natural choice may be \textit{doit}, which operates in Python and is easy to use for
anyone familiar with the language. Another benefit of this system is that one can use
Python functions as processes, making it possible to easily transfer data from one process
to the other via memory without the need to write and read to disk. In order to make a
workflow portable, developers have to provide additional resources that allow users
to prepare their environment such that all software dependencies are met, prior to the
workflow execution.

To create portable workflows more easily, convenient tools are \textit{Nextflow} or \textit{Snakemake},
where one can specify the compute environment in terms of a conda environment file or a container image on a per-process basis. They require to learn a new 
domain-specific language, however, our assessment is that it is easy to get started as only little syntax has to be learned in order to get a first workflow 
running.

The strengths of \textit{AiiDA} are the native support for distributing the workload
on different (registered) machines, the comprehensive provenance tracking, and also
the possibility to transfer data among processes without the creation of intermediate files.

\textit{CWL} has the benefit of being a language standard rather than a specific tool
maintained by a dedicated group of developers. This has led to a variety of tooling
developed by the community as \eg editors for visualizing and modifying workflows
with a GUI\@. Moreover, the workflow description states the version
of the standard in which it is written, such that any interpreter supporting this
standard should execute it properly, which reduces the problem of version pinning on
the level of the workflow interpreter.

Especially for larger workflows composed of processes that are still under development,
and are thus changing over time, it may be useful to rely on tools that allow to define
the process interfaces by means of strongly-typed arguments. This can help to detect
errors early on, \eg by static type checkers. \textit{CWL} and \textit{AiiDA} support
the definition of strongly-typed process interfaces.
The rich set of options and features of these tools
make them more difficult to learn, but at the same time expose a large number of possibilities.

\section{Outlook}
This overview is not meant to be static, but we plan to continue the documentation
online in the git repository~\cite{NFDI4Ing_Scientific_Workflow_2022} that contains
the implementation of the simple use case.
This allows us to take into consideration other WfMSs in the future, and to extend the documentation accordingly.
In particular, we would like to make the repository a community effort allowing others to contribute either by modifications of the existing tools or adding new WfMSs.
All of our workflow implementations are continuously and automatically tested using GitHub Actions \url{https://github.com/BAMresearch/NFDI4IngScientificWorkflowRequirements/actions}, which may act as an additional source of documentation on how to launch the workflows.

One of the challenges we have identified is the use of container technology
in the HPC environment.
In most cases, the way users should interact with such a system is through a
module system provided by the system administrators.
The module system allows to control the software environment (versioning, compilers)
in a precise manner, but the user is limited to the provided software stack.
For specific applications, self-written code can be compiled using the available
development environment and subsequently run on the system, which is currently 
the state of the art in using HPC systems.
However, this breaks the portability of the workflow.

Container technology, employing the \enquote{build once and run anywhere} 
concept, seems to be a promising solution to this problem.
Ideally, one would like to be able to run the container application on
the HPC system, just as any other MPI-distributed application.
Unfortunately, there are a number of problems entailed with this approach.

When building the container, great care must be taken with regard to the
MPI configuration, such that it can be run successfully across several nodes.
Another issue is the configuration of Infiniband drivers.
The container has to be build according to the specifics of the HPC system 
that is targeted for execution.
From the perspective of the user, this entails a large difficulty, and we think that
further work needs to be done to find solutions which enable non-experts in
container technology to execute containerized applications successfully in
an HPC environment.

Furthermore, challenges related to the joint development of workflows became
apparent.
In this regard, strongly-typed interfaces are required in order to minimize errors
and transparently and clearly communicate the metadata (inputs, outputs) associated with a process in the workflow.
This is recommended both for single parameters, but it would be also great to extend that idea to files - not only defining the file type which is already possible within \textit{CWL} - but potentially allowing a type checking of the complete data structure within the file.
However, based on our experience with the selected tools, these interfaces and
their benefits come at the cost of some form of
plugin or wrapper around the software that is to be executed, thus possibly
limiting the functionality of the wrapped tool.
This means there is a trade-off between easy authoring of the workflow definition
(\eg easily executing any shell command) and implementation overhead
for the sake of well-defined interfaces.

Another aspect is how the workflow logic can be communicated efficiently.
Although each of the tools allows to generate a graph of the workflow, 
the dependencies between processes can only be visualized for an executable
implementation of the workflow, which most likely does not exist in early
stages of the project where it is needed the most.

An important aspect is the documentation of the workflow results and how they have been obtained. Most tools offer an option to export the data provenance graph, however it would be great to define a general standard supported by all tools as e.g. provided by \textit{CWLProv}~\cite{Khan2019}.

%\backmatter

\section*{Acknowledgments}
The authors would like to thank the Federal Government and the Heads of Government of the Länder, as well as the Joint Science
Conference (GWK), for their funding and support within the framework of the NFDI4Ing consortium. Funded by the German Research
Foundation (DFG) - project number 442146713.
Moreover, we would like to thank Sebastiaan P. Huber, Michael R. Crusoe,
Eduardo Schettino, Ricardo Wurmus, Paolo Di Tommaso and Johannes Köster
for their valuable remarks and comments on an earlier version of this article
and the workflow implementations.

\subsection*{Author contributions}
\textbf{Philipp Diercks:} Investigation; methodology; software; writing - original draft; writing - review and editing.
\textbf{Dennis Gläser:} Investigation; methodology; software; writing - original draft; writing - review and editing.
\textbf{Ontje Lünsdorf:} Investigation (supporting); software; writing - original draft (supporting).
\textbf{Michael Selzer:} Writing - review and editing (supporting).
\textbf{Bernd Flemisch:} Conceptualization (supporting); Funding acquisition; Project administration; Writing - review and editing.
\textbf{Jörg F. Unger:} Conceptualization (lead); Funding acquisition; Project administration; Writing - original draft (supporting); Writing - review and editing.

\subsection*{Financial disclosure}

None reported.

\subsection*{Conflict of interest}

The authors declare no potential conflict of interests.

\bibliography{literature,software}%

\begin{thebibliography}{10}
\providecommand \doibase [0]{http://dx.doi.org/}%

\bibitem{FAIR2016}
Wilkinson MD, Dumontier M, Aalbersberg IJ, et al. The {FAIR} Guiding Principles
  for scientific data management and stewardship. {\it Sci Data} 2016\string;
  3(1).
\newblock \href {\doibase 10.1038/sdata.2016.18} {doi: 10.1038/sdata.2016.18}

\bibitem{MonsEtAl2020}
Mons B, Schultes E, Liu F, Jacobsen A. The {FAIR} Principles: {First}
  Generation Implementation Choices and Challenges. {\it Data Intellegence}
  2020\string; 2(1-2)\string: 1-9.
\newblock \href {\doibase 10.1162/dint\_e\_00023} {doi: 10.1162/dint\_e\_00023}

\bibitem{EOSC}
{Directorate-General for Research and Innovation (European Commission)} . First
  report and recommendations of the Commission high level expert group on the
  European Open Science Cloud. Available at \url{https://op.europa.eu/s/wGAL};
  2016

\bibitem{fair4rs}
Chue~Hong NP, Katz DS, Barker M, et al. {FAIR Principles for Research Software
  ({FAIR4RS} Principles)}. \url{https://doi.org/10.15497/RDA00068};  2022

\bibitem{GobleEtAl2020}
Goble C, Cohen-Boulakia S, Soiland-Reyes S, et al. {FAIR Computational
  Workflows}. {\it Data Intelligence} 2020\string; 2(1-2)\string: 108-121.
\newblock \href {\doibase 10.1162/dint\_a\_00033} {doi: 10.1162/dint\_a\_00033}

\bibitem{NFDI4Ing_Scientific_Workflow_2022}
Diercks P, Gläser D, Unger JF, Flemisch B. {NFDI4Ing Scientific Workflow
  Requirements}.;  2022.
\newblock
  \url{https://github.com/BAMresearch/NFDI4IngScientificWorkflowRequirements}.

\bibitem{griem2022kadistudio}
Griem L, Zschumme P, Laqua M, et al. KadiStudio: FAIR Modelling of Scientific
  Research Processes. {\it Data Science Journal} 2022\string; 21(1).

\bibitem{Snakemake2021}
Mölder F, Jablonski KP, Letcher B, et al. Sustainable data analysis with
  Snakemake. {\it F1000Res} 2021\string; 10\string: 33.
\newblock \href {\doibase 10.12688/f1000research.29032.2} {doi:
  10.12688/f1000research.29032.2}

\bibitem{galaxy}
Afgan E, Baker D, Batut B, et al. {The Galaxy platform for accessible,
  reproducible and collaborative biomedical analyses: 2018 update}. {\it
  Nucleic Acids Research} 2018\string; 46(W1)\string: W537-W544.
\newblock \href {\doibase 10.1093/nar/gky379} {doi: 10.1093/nar/gky379}

\bibitem{knime}
Berthold MR, Cebron N, Dill F, et al. KNIME: The Konstanz Information Miner.
  In: Data Analysis , Machine Learning and Applications : Proceedings of the
  31st Annual Conference of the Gesellschaft für Klassifikation e. V.,
  Albert-Ludwigs-Universität Freiburg, March 7-9 , 2007. Springer; 2007; New
  York.

\bibitem{pegasus}
Deelman E, Vahi K, Juve G, et al. Pegasus, a workflow management system for
  science automation. {\it Future Generation Computer Systems} 2015\string;
  46\string: 17-35.
\newblock \href {\doibase 10.1016/j.future.2014.10.008} {doi:
  10.1016/j.future.2014.10.008}

\bibitem{HuberEtal2020}
Huber SP, Zoupanos S, Uhrin M, et al. {AiiDA} 1.0, a scalable computational
  infrastructure for automated reproducible workflows and data provenance. {\it
  Sci Data} 2020\string; 7(1).
\newblock \href {\doibase 10.1038/s41597-020-00638-4} {doi:
  10.1038/s41597-020-00638-4}

\bibitem{UhrinEtAl2021}
Uhrin M, Huber SP, Yu J, Marzari N, Pizzi G. Workflows in {AiiDA:}
  {Engineering} a high-throughput, event-based engine for robust and modular
  computational workflows. {\it Nato. Sc. S. Ss. Iii. C. S.} 2021\string;
  187\string: 110086.
\newblock \href {\doibase 10.1016/j.commatsci.2020.110086} {doi:
  10.1016/j.commatsci.2020.110086}

\bibitem{pydoit}
Schettino EN. {pydoit/doit: {Task} management \& automation tool (python)}.
  \url {https://doi.org/10.5281/zenodo.4892136};  2021

\bibitem{balsam_preprint}
Salim MA, Uram TD, Childers JT, Balaprakash P, Vishwanath V, Papka ME. Balsam:
  Automated Scheduling and Execution of Dynamic, Data-Intensive HPC Workflows.
  \url{https://arxiv.org/abs/1909.08704};  2019

\bibitem{fireworks}
Jain A, Ong SP, Chen W, et al. {FireWorks:} {A} dynamic workflow system
  designed for high-throughput applications. {\it Concurrency Computat.: Pract.
  Exper.} 2015\string; 27(17)\string: 5037-5059.
\newblock \href {\doibase 10.1002/cpe.3505} {doi: 10.1002/cpe.3505}

\bibitem{scipipe}
Lampa S, Dahlö M, Alvarsson J, Spjuth O. {SciPipe: A workflow library for
  agile development of complex and dynamic bioinformatics pipelines}. {\it
  GigaScience} 2019\string; 8(5).
\newblock \href {\doibase 10.1093/gigascience/giz044} {doi:
  10.1093/gigascience/giz044}

\bibitem{Nextflow2017}
Di~Tommaso P, Chatzou M, Floden EW, Barja PP, Palumbo E, Notredame C. Nextflow
  enables reproducible computational workflows. {\it Nat Biotechnol}
  2017\string; 35(4)\string: 316-319.
\newblock \href {\doibase 10.1038/nbt.3820} {doi: 10.1038/nbt.3820}

\bibitem{Snakemake2012}
Köster J, Rahmann S. Snakemake---a scalable bioinformatics workflow engine.
  {\it Method. Biochem. Anal.} 2018\string; 34(20)\string: 3600-3600.
\newblock \href {\doibase 10.1093/bioinformatics/bty350} {doi:
  10.1093/bioinformatics/bty350}

\bibitem{bpipe}
Sadedin SP, Pope B, Oshlack A. Bpipe: {A} tool for running and managing
  bioinformatics pipelines. {\it Method. Biochem. Anal.} 2012\string;
  28(11)\string: 1525-1526.
\newblock \href {\doibase 10.1093/bioinformatics/bts167} {doi:
  10.1093/bioinformatics/bts167}

\bibitem{gwl2022}
Wurmus R, others . {GUIX Workflow Language}. \url{https://guixwl.org};  2022.

\bibitem{Ewels2016}
Ewels P, Krueger F, K{\"{a}}ller M, Andrews S. Cluster Flow: A user-friendly
  bioinformatics workflow tool [version 2; referees: 3 approved].. {\it
  F1000Research} 2016\string; 5\string: 2824.
\newblock \href {\doibase 10.12688/f1000research.10335.2} {doi:
  10.12688/f1000research.10335.2}

\bibitem{popper}
Jimenez I, Sevilla M, Watkins N, et al. The Popper Convention: {Making}
  Reproducible Systems Evaluation Practical. In: 2017 IEEE International
  Parallel and Distributed Processing Symposium Workshops (IPDPSW). IEEE;
  2017\string: 1561-1570

\bibitem{yaml}
Ben-Kiki O, Evans C, Net dI. YAML Ain’t Markup Language (YAML) version 1.2.
  Accessed: 2022-08-31;  2021.
\newblock \url{https://yaml.org/spec/1.2.2/}.

\bibitem{cwl_project_june_2022}
Crusoe MR, Abeln S, Iosup A, et al. Methods included. {\it Commun. ACM}
  2022\string; 65(6)\string: 54-63.
\newblock \href {\doibase 10.1145/3486897} {doi: 10.1145/3486897}

\bibitem{cromwell}
Voss K, Auwera GVD, Gentry J. Full-stack genomics pipelining with GATK4 + WDL +
  Cromwell [version 1; not peer reviewed].;  2017.
\newblock \url{https://f1000research.com/slides/6-1381}

\bibitem{toil}
Vivian J, Rao AA, Nothaft FA, et al. Toil enables reproducible, open source,
  big biomedical data analyses. {\it Nat Biotechnol} 2017\string; 35(4)\string:
  314-316.
\newblock \href {\doibase 10.1038/nbt.3772} {doi: 10.1038/nbt.3772}

\bibitem{tibanna}
Lee S, Johnson J, Vitzthum C, Kırlı K, Alver BH, Park PJ. Tibanna: {Software}
  for scalable execution of portable pipelines on the cloud. {\it Method.
  Biochem. Anal.} 2019\string; 35(21)\string: 4424-4426.
\newblock \href {\doibase 10.1093/bioinformatics/btz379} {doi:
  10.1093/bioinformatics/btz379}

\bibitem{GeuzaineRemacle2009}
Geuzaine C, Remacle JF. Gmsh: {A} 3-D finite element mesh generator with
  built-in pre- and post-processing facilities. {\it Int. J. Numer. Meth.
  Engng.} 2009\string; 79(11)\string: 1309-1331.
\newblock \href {\doibase 10.1002/nme.2579} {doi: 10.1002/nme.2579}

\bibitem{AlnaesEtal2015}
Alnaes M, Blechta J, Hake J, et al. The {FEniCS} Project Version 1.5. {\it
  Archive of Numerical Software} 2015\string; 3.
\newblock \href {\doibase 10.11588/ans.2015.100.20553} {doi:
  10.11588/ans.2015.100.20553}

\bibitem{meshio534}
Schlömer N. meshio: {Tools} for mesh files. \url
  {https://doi.org/10.5281/zenodo.6346837};  2022

\bibitem{Schroeder2006}
Schroeder W, Martin K, Lorensen B, {Kitware} . {\it The visualization toolkit :
  an object-oriented approach to {3D} graphics}.
\newblock Kitware.
\newblock 4~ed. 2006.

\bibitem{Ahrens2005ParaViewAE}
Ahrens J, Geveci B, Law C. {ParaView:} {An} End-User Tool for Large-Data
  Visualization. In: Elsevier; 2005.

\bibitem{tectonic}
Williams P, Contributors . The Tectonic Typesetting System. \url
  {https://tectonic-typesetting.github.io/en-US/};  2022.
\newblock Accessed: 2022-06-02.

\bibitem{Huber_aiida-shell_2022}
Huber SP. {aiida-shell}.;  2022.
\newblock \url{https://github.com/sphuber/aiida-shell}.

\bibitem{Amstutz2016}
Amstutz P, Crusoe MR, Tijanić N, et al. {Common Workflow Language, v1.0}.;
  2016.
\newblock \url{https://doi.org/10.6084/m9.figshare.3115156.v2}

\bibitem{Colonnelli_StreamFlow_cross-breeding_cloud_2021}
Colonnelli I, Cantalupo B, Merelli I, Aldinucci M. {StreamFlow: cross-breeding
  cloud with HPC}. {\it IEEE Transactions on Emerging Topics in Computing}
  2021\string; 9(4)\string: 1723--1737.
\newblock \href {\doibase 10.1109/TETC.2020.3019202} {doi:
  10.1109/TETC.2020.3019202}

\bibitem{slurm_2003}
Yoo AB, Jette MA, Grondona M. SLURM: Simple Linux Utility for Resource
  Management. In:  Feitelson D, Rudolph L, Schwiegelshohn U. \kern-2pt, eds.
  {\it Job Scheduling Strategies for Parallel Processing}Springer Berlin
  Heidelberg; 2003; Berlin, Heidelberg\string: 44--60.

\bibitem{Courtes2013}
Courtès L. Functional Package Management with Guix. {\it {European Lisp
  Symposium}} 2013.
\newblock \href {\doibase 10.48550/ARXIV.1305.4584} {doi:
  10.48550/ARXIV.1305.4584}

\bibitem{Courtes2022}
Courtès L. Building a Secure Software Supply Chain with {GNU} Guix. {\it The
  Art, Science, and Engineering of Programming} 2022\string; 7(1).
\newblock \href {\doibase 10.22152/programming-journal.org/2023/7/1} {doi:
  10.22152/programming-journal.org/2023/7/1}

\bibitem{akhlagi2022}
Akhlaghi M, Infante-Sainz R, Roukema BF, Khellat M, Valls-Gabaud D, Baena-Galle
  R. Toward Long-Term and Archivable Reproducibility. {\it Computing in Science
  \& Engineering} 2021\string; 23(3)\string: 82--91.
\newblock \href {\doibase 10.1109/mcse.2021.3072860} {doi:
  10.1109/mcse.2021.3072860}

\bibitem{Sperber2009}
Sperber M, Dybvig RK, Flatt M, Van~Straaten A, Findler R, Matthews J. Revised6
  Report on the Algorithmic Language Scheme. {\it J. Funct. Program.}
  2009\string; 19(S1)\string: 1.
\newblock \href {\doibase 10.1017/s0956796809990074} {doi:
  10.1017/s0956796809990074}

\bibitem{guile2022}
Wingo A, others . {GNU Guile}. \url{https://www.gnu.org/software/guile/};
  2022.

\bibitem{wisp2015}
Babenhauserheide A. SRFI 119: wisp: simpler indentation-sensitive scheme.
  \url{https://srfi.schemers.org/srfi-119/};  2015.

\bibitem{Khan2019}
Khan FZ, Soiland-Reyes S, Sinnott RO, Lonie A, Goble C, Crusoe MR. Sharing
  interoperable workflow provenance: A review of best practices and their
  practical application in CWLProv. {\it GigaScience} 2019\string; 8(11).
\newblock \href {\doibase 10.1093/gigascience/giz095} {doi:
  10.1093/gigascience/giz095}

\end{thebibliography}

\end{document}